\documentclass[aps,pre,reprint,groupedaddress,longbibliography]{revtex4-1}
\usepackage{times}
\usepackage{chemarrow} 
\usepackage{amsmath,amssymb,mathrsfs}
\usepackage{accents}
\usepackage{mathtools}
\usepackage{graphicx,epsfig}
\usepackage{xcolor}
\usepackage{bbm}
\usepackage{extarrows,chemarrow,xypic}
\usepackage{chemarrow}
\usepackage{hyperref}
\usepackage{microtype}
\DisableLigatures[f]{encoding = *, family = *}
\usepackage[none]{hyphenat}
\usepackage{lipsum}
\def\rd{{\rm d}}

\def\vGamma{\mbox{\boldmath$\Gamma$}}

\begin{document}


\title{Thermodynamic Behavior of Statistical Event Counting in Time:\\  Independent and Correlated Measurements}



\author{Hong Qian}

\affiliation{Department of Applied Mathematics, University of Washington,
Seattle, WA 98195-3925, USA}



\begin{abstract}
We introduce an entropy analysis of time series,
repeated measurements of statistical observables, based on an Eulerian homogeneous degree-one entropy function $\Phi(t,n)$ of time $t$ and number of events $n$.  The duality of $\Phi$, in terms of conjugate variables $\eta=-\Phi'_t$ and $\mu=\Phi'_n$,  yields an ``equation of state'' (EoS) in differential form that resembles the Gibbs-Duhem relation in classical thermodynamics: $t\rd\eta-n\rd\mu = 0$.  For simple Poisson counting with rate $r$, $\eta=r(e^{\mu}-1)$.  The conjugate variable $\eta$ is then identified as being equal to the Hamiltonian function in a Hamilton-Jacobi equation for $\Phi(t,n)$. Applying the same logic to the entropy function of time correlated events yields a Hamiltonian as the principal eigenvalue of a matrix.  For time reversible case it is the sum of a symmetric Markovian part $\sqrt{\pi_i}q_{ij}/\sqrt{\pi_j}$ and the conjugate variables $\mu_i\delta_{ij}$.  The eigenvector, as a posterior to the naive counting measure used as the prior, suggests a set of intrinsic characteristics of Markov states.

\end{abstract}


\maketitle




\section{Introduction}

	Recent studies \cite{lu-qian-20,cyq-21} on (i) the probabilistic foundation of Hill's nanothermodynamics (1963) \cite{hill-book,bedeaux}, (ii) the duality symmetry hidden in Gibbs' statistical mechanics and thermochemistry \cite{bennaim,schellman}, and (iii) the revisitation of the Gibbs and the Shannon entropies, all have shown central importance of the large deviations theory \cite{dembo,rassoul-agha,ge-qian-ijmpb,vulpiani} for the asymptotic, entropic description of statistical observables, such as empirical sample mean values and counting frequencies, as well as the key role played by conjugate variables in establishing the concept of an equation of state (EoS). It was revealed that the Gibbs entropy as a function of mean internal energy is the Legendre-Fenchel dual to the Massieu potential as a function of inverse temperature, and the Helmhotz free energy as a function of a set of energies of states in $k_BT$ units, $\{u_i\}$, is the Legendre-Fenchel dual to the Shannon-Sanov relative entropy as a function of empirical counting frequencies $\{\nu_i\}$.  Identifying proper potential functions and their  corresponding independent variabels is paramount in classical thermodynamic analysis as well as in Gibbs' theory of statistical ensembles. 

	Statistical thermodynamics is a theory about emergent phenomenon \cite{chibbaro}.  The recent mathematical development \cite{lu-qian-20,cyq-21} in fact had been captured presciently by P. W. Anderson in 1972 \cite{pwanderson},
\begin{quote}
{\em Starting with the fundamental laws and a computer, we would have to do two [impossible] (unthinkable) things | solve a problem with infinitely many bodies, and then apply the result to a finite system | before we synthesized [this] (an emergent) behavior.
}
\end{quote}
Here ``solving a problem with infinitely many bodies'' is now understood as taking a mathematical limit w.r.t. a parameter in a model. In many-body physics, this has been routinely the {\em thermodynamic limit}.  For a probabilistic description of a phenomenon, what Anderson has not delineated are in fact {\em three} levels of descriptions: a law of large numbers (LLN) that determines the macroscopic behavior, a central limit theorem (CLT) describing the fluctuations, and the large deviations theory (LDT) that provides an entropic description of the macroscopic system in terms of an optimization problem together with an objective function \cite{dembo,vulpiani}.  A large deviation rate function (LDRF), 
as the entropy, actually embodies the LLN as its global minimum and the CLT as the quadratic Hessian structure nearby.  But it is the entropy  function itself that yields a variational principle that represents the fundamental tenet of thermodynamics.

\begin{figure}
\[
\includegraphics[scale=0.34]{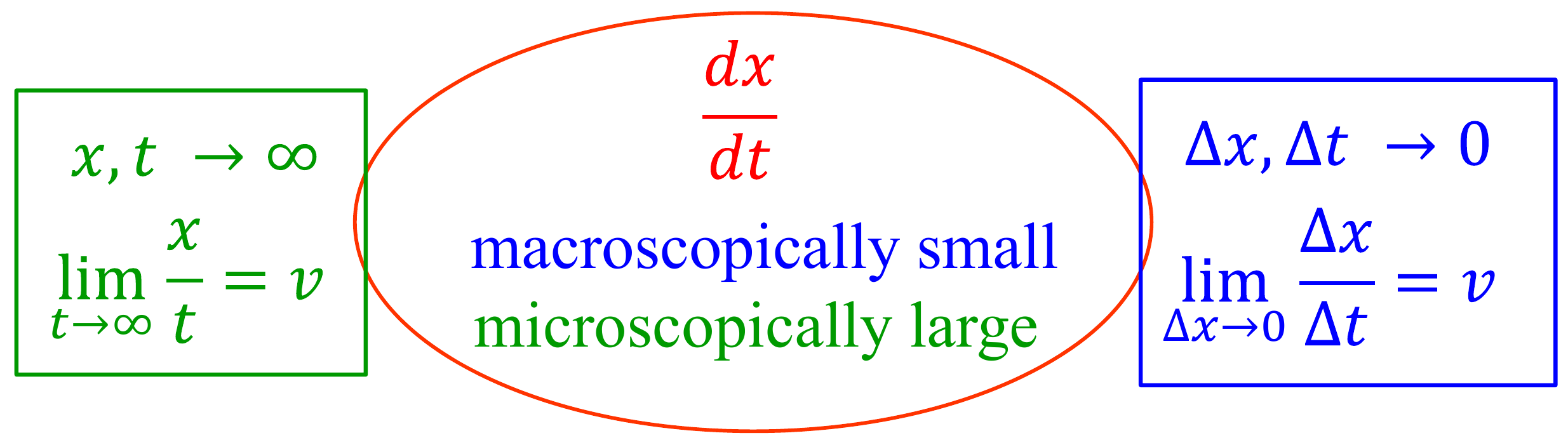}
\]
\caption{The concept of instantaneous rate of change, $v(t)$, can be understood under two very different but equally fundamental mathematical suppositions of the real world: (i) From top down with a deterministic, smooth description at a {\em macroscopic time scale}, as the limit of the ratio of infinitesimal $\Delta x(t)$ and $\Delta t$. (ii) From bottom up with a stochastic description at a {\em microscopic time scale}, as the limit of accumulation of a large number of microscopic events $x(t)$ that occur randomly along time $t$.  A thermodynamic, entropy analysis naturally arises from (ii) following the law of large numbers (LLN) and the large deviations theory (LDT). The concept of entropy also emerges in (i), via chaos theory using the idea of {\em Markov partition} \cite{dorfman-book} and related thermodynamic formalism \cite{ruelle-book}.
}
\label{figure1} 
\end{figure}

	In parallel with thermodynamic limit of macroscopic size, another important applications of LDT is the statistical counting of number of probabilistic events on ``a microscopic time scale'', $N(t)$, which in the limit of large numbers of measurements yields the concept of instantaneous rate $r=N(t)/t$, see Fig. \ref{figure1}.  The very existence of the $r$ in ``macroscopic time scale'' implies that $N(t)\sim O(t)$ as $t\to\infty$.  Therefore one has 
\begin{equation}
\label{elike-eq}
     \Phi(t,n)\equiv \log \Pr\big\{ N(t) = n \big\} = 
           \Phi_0(t,n) + o(t),
\end{equation}
where $\Phi_0(t,n)$ is an entropy like function that is Eulerian homogeneous degree-one (Ehd-1) w.r.t. the ``extenstive'' variables $n$ and $t$.  Introducing $\nu=n/t$, LDT states that $\Phi(t,\nu t)/t\to \Phi_0(t,\nu t)/t \equiv -\varphi(\nu)$ as $t\to\infty$, where $\varphi(\nu)$ is the LDRF in the probability theory applied to 
sample statistics of big time-series data: The Ehd-1 entropy function $\Phi_0$ in physics and LDRF $\varphi$ in mathematics are indentical concept.

	Following the logic of classical thermodynamics as presented by H. B. Callen in \cite{callen} and further developed recently in \cite{lu-qian-20,cyq-21}, a macroscopic thermodynamics like analysis for the $\Phi(t,n)$ in (\ref{elike-eq}) can be carried out.  This yields an {\em equation of state} (EoS) among two conjugate intensive variables corresponding to time ($t$) and statistical counting ($n$).  Hill's thermodynamics of small systems \cite{hill-book}, now dubbed {\em nanothermodynamics} \cite{bedeaux}, entails a further analysis of (\ref{elike-eq}) by including the neglected higher order, sub-extensive (subadditive) terms that breaks the entropy function being Ehd-1 (see Eq. \ref{equation16} below).  One of the important examples of nanothermodynamics with sub-extensive effect is from the surface tension of a droplet  \cite{kjelstrup}.

	Fig. \ref{figure1} captures the relationship between formulating the concept of instantaneous rate according to the traditional way, as the ratio of infinitesimal changes with $\Delta t\to 0$, and the stochastic large-numbers limit in which both $x$ and $t$ are tending to infinity.  
One of the best examples for the latter is the concept of instantaneous rate of a chemical reaction \cite{ge-qian-2016}. 

	In the present paper, we first use the Poisson process of identical, independent random events to illustrate the essential idea of a ``thermodynamics in time''. We apply this idea toward a novel interpretation of classical mechanics' Lagrangian and Hamiltonian formalisms.  We then consider time-correlated events represented by stationary Markov processes, and develop a nanothermodynamic treatment for the subadditive entropy in the Poisson case.  Sec. \ref{sec:discuss} discusses a wide range of implications from the entropy analysis of time.

\section{Thermodynamics of time and independent events}

Repeated measurement on a time-invariant system is one of the most important activities of modern research.  For time stationary stochastic phenomena the notion of {\em rate} for random events is a fundamental concept.  It follows then a ``thermodynamics'' for a large number of random events over a long time, corresponding to the thermodynamics of a macroscopic system.  For relatively shorter time, there should also be a corresponding thermodynamics of small systems \cite{hill-book}; the latter will be carried out in Sec. \ref{sec:iv}.   

	We shall use the simplest Poisson process to illustrate our essential idea through explicit calculations. The probability distribution for the Poisson process with a constant rate $r$ is 
\begin{equation}
        \Pr\big\{ N(t) = n\big\} = \frac{(r t)^n}{n!}e^{-r t},
\end{equation}
which gives an entropy like quantity in asymptotic orders
\begin{subequations}
\label{equation16}
\begin{eqnarray}
     \Phi(t,n) &=&  -\left\{ n\log\left(\frac{n}{r t}\right) - n + r t \right\} 
\\
	&-& \left[ \frac{n+1}{2}\log(n+1)-\frac{n}{2}\log n \right] + O(1),
   \hspace{0.7cm}
\end{eqnarray}
\end{subequations}
as $n, t\to\infty$ and $n\sim O(t)$.  Noting that the term inside $\{\cdots\}$ is an Ehd-1 function.  The term inside $[\cdots]$, which is on the order of $O(\log n)$, breaks Euler's homogeneous property.

\subsection{Macroscopic thermodynamic treatment}

We first consider only the $\{\cdots\}$ term in (\ref{equation16}a):
$\Phi_0(t,n)=-n\log(n/rt)+n-rt$. Then $\Phi_0$ has two {\em partial Legendre-Fenchel transforms} with conjugate variables $\eta$ and $\mu$ corresponding to $t$ and $n$:
\begin{subequations}
\label{equation17}
\begin{eqnarray}
	       && \eta = \left(\frac{\partial \Phi_0}{\partial t}\right)_n
            = \frac{n}{t} - r,
\\
	 && \tilde{\Psi}_1(\eta,n) = \Phi_0 - \eta t
          = n\log\left(\frac{r}{r +\eta}\right),
\\
	 && \mu =  -\left(\frac{\partial \Phi_0}{\partial n}\right)_t
            = \log\left(\frac{n}{r t}\right),
\\
	&& \tilde{\Psi}_2(t,\mu) =  \Phi_0 +\mu n 
           = r t\big(  e^{\mu} -1\big).
\end{eqnarray}
\end{subequations}
One therefore has a ``fundamental thermodynamic relation'':
\begin{equation}
             \rd\Phi_0 = \left(\frac{n}{t}- r\right)\rd t 
              - \log\left(\frac{n}{r t}\right) \rd n = \eta \rd t - \mu \rd n,
\label{equation18}
\end{equation}
and a Gibbs-Duhem equation like relation:
\begin{equation}
   \Phi_0 =  \eta t - \mu n,
\end{equation}
with its differential form:
\begin{equation}
          t\rd\eta - n\rd\mu = 0\  \text{ or }  \
                \frac{t}{n} = \frac{\rd\mu}{\rd\eta}.
\label{gde}
\end{equation}
Eq. \ref{gde} implies an EoS among $\eta$ and $\nu$.  The form specific for the Poisson counting process is obtained by setting either the homogeneous $\tilde{\Psi}_1(\eta,n)=n\mu$, or $\tilde{\Psi}_2(t,\mu)=t\eta$:
\begin{equation}
\label{eos}
             \mu = \log\left(\frac{r +\eta}{r}\right) \
 \text{ or } \      \eta = r\big(e^{\mu}-1\big).
\end{equation}
These two relations are the same as expected.  It satisfies Eq. \ref{gde} and it has $\eta=0$ if and only if $\mu=0$.

In general, the macroscopic EoS is obtained as the root of the singular Legendre-Fenchel transform of homogeneous functions $\tilde{\Psi}_1(\eta,n)$ or $\tilde{\Psi}_2(t,\mu)$:
\begin{equation}
   \frac{\partial\tilde{\Psi}_1(\eta,n)}{\partial n} = 
   \frac{\tilde{\Psi}_1}{n}=-\mu   \ \text{ or } \
  \frac{\partial\tilde{\Psi}_2(t,\mu)}{\partial t} = \frac{\tilde{\Psi}_2}{t} = \eta. 
\label{EoS-1}
\end{equation}
A similar idea was behind the method of divergent generalized partition function, independently developed by M. Kac {\it et. al.} \cite{kac} and by S. Lifson \cite{lifson} in 1960s. 

	Note also that (\ref{eos}) can be obtained directly from combining (\ref{equation17}a) and (\ref{equation17}c) by eliminating $(n/t)$.
A geometric statement for the existence of an EoS is that the variable transformation from extensive $(t,n)$ to their conjugate variables $(\eta,\mu)$ is non-invertible \cite{arnold,ruppeiner}.

	The relation between the entropy analysis based on Ehd-1 $\Phi_0$
and the LDRF is simple and straightforward: $\varphi(\nu)=-\Phi_0(t,\nu t)/t$ in which $\nu=n/t$, the intensive stochastic rate which tends to $r$ as $t\to\infty$.  Through chain rules it is easy to see that the EoS in Eq. \ref{EoS-1} is a simple incarnation of
\begin{equation}
    \eta = -\sup_{\nu} \big\{ \mu\nu-\varphi(\nu)\big\},
\label{e9}
\end{equation}
the rhs of which is the Legendre-Fenchel transform (LFT) of $\varphi(\nu)$; it is a function of $\mu$.  The time $t$ has a very special place in LDT: It is a parameter.  Since $\Phi_0$ is a homogeneous function of $t$ and $n$, its conjugate variable $\eta$ satisfying (\ref{e9}) is immediately true.  Then it follows that 
\begin{quote}
{\em The equation of state for time and random events is that the conjugate variable of time being equal to the Legendre-Fenchel transform of the large deviation rate function.}
\end{quote}

\subsection{The meaning of thermodynamics of time and counting events} 

While the stochastic counting model according to Poisson process is extremely simple, the logic in the previous section is clear and general: There is a definite mathematical relation among the conjugate variables, with its functional form to be determined specifically case by case: This is the significance of the EoS in classical thermodynamics of equilibrium matters.  It is important to recall that the same EoS is a differential equation for the entropy function $\Phi_0(t,n)$:
\begin{equation}
  \frac{\partial\Phi_0}{\partial t} = r\left[
         \exp\left(-\frac{\partial\Phi_0}{\partial n}\right) - 1\right].
\label{hje0}
\end{equation}

	Observing the EoS in Eq. \ref{hje0} and the relation $\rd\eta = r \rd \mu$ in Eq. \ref{gde}, it is tempting to interpret classical mechanical equation of motion as a relation between mechanical energy and momentum, the respective conjugate variables to time and velocity: $\rd E =v F \rd t = mv\rd v = v \rd p$ where $E$, $v$, and $p$ are mechanical energy, velocity, and momentum of a point mass $m$.  One also notices that the differential equation in (\ref{hje0}) is the type of nonlinear partial differential equations called a Hamilton-Jacobi equation (HJE), with the rhs being a Hamiltonian function, whose specific form is to be determined case-by-case.  Thus, it is not groundless to suggest that the mechanical law of motions could be understood alternatively as an EoS for time and statistical counting of random displacements in space.

\subsection{A data science perspective on mechanical movements}

	To further substantiate the above hypothesis, we now generalize the earlier result for Poisson counting to an arbitrary time series data on mechanical movements in the form of time intervals $(\Delta t)_j$ and corresponding ``displacement'' $(\Delta x)_j$, $j=1,2,\cdots,J$.  Denoting 
\begin{equation}
         X =  \sum_{j=1}^J (\Delta x)_j, \ 
         t  = \sum_{j=1}^J (\Delta t)_j,
\end{equation}
then with increasing time, one has the law of large numbers $X/t \to v$, the velocity, as $t\to\infty$.  Let the Ehd-1 entropy function of $X$ and $t$ as $\Phi_0(t,X)$, then
\begin{eqnarray}
   \rd \Phi_0(t,X) &=&  \left(\frac{\partial\Phi_0}{\partial X}\right)
             \rd X + \left(\frac{\partial\Phi_0}{\partial t}\right) \rd t 
\nonumber\\
           &=& p(X,t) \rd X - H(X,t) \rd t,  
\label{eq35}
\end{eqnarray}
in which we have introduced the conjugate variables to $X$ and $t$ as
$p$ and $-H$.  While $X$, $t$, and $\Phi_0$ are all ``extensive variables'', the $p$ and $H$ are intensive variables.  This implies they both are actually functions of $v=X/t$ only.  Then for Ehd-1 entropy function $\Phi_0(t,X)$:
\begin{subequations}
\label{eq36}
\begin{eqnarray}
    && \Phi_0(t,X) = p(v)X - H(v)t,
\\
	&& H = p\frac{X}{t} - \frac{\Phi_0(t,X)}{t}
            = p\frac{X}{t} - \Phi_0\left(\tfrac{X}{t},1\right),
      \hspace{0.8cm}
\\
	&& H = pv + L(v),
\end{eqnarray}
in which we have denoted $\Phi_0(1,X/t)$ by $L(v)$, the LDRF, and 
$\rd L/\rd v = p(v)$.  As the Gibbs-Duhem equation in classical thermodynamics, combining Eqs. \ref{eq35} and
\ref{eq36}a one has:
\begin{equation}
            \rd H - v\rd p = 0.
\end{equation}
\end{subequations}
$L(v)$ as a function of $v$ and $H\big(v(p)\big)$ as a function of $p$ are Legendre transforms of each other; they could be identified as Lagrangian and Hamiltonian functions.  The former is the LDRF for the empirical mean displacement $v=X/t$ as $t\to\infty$, and the latter is the conjugate variable to time $t$, in terms of the entropy function $\Phi_0$, with time $t$ as a special variable:
\begin{equation}
     \frac{\partial\Phi_0}{\partial t} - H\left[v\left(\frac{\partial\Phi_0}{\partial X}\right)\right] = 0,
\label{HJE}
\end{equation}
in which $\Phi_0$ can be identified with the action in a stationary action principle.

	The Lagrangian function $L(\dot{x})$ is independent of space $x$ since we have assumed a translational invariance in the simple counting problem.  A field theory formulation is required when space $x$ is explicitly considered; this can be accomplished via Feynman's formalism to space-time physics \cite{feynman}.  As a LDRF, the $L(\dot{x})$ is a divergence function on the tangent space of a manifold.  According to information geometry, it offers both a local Riemannian metric as well as a nonlinear connection between local charts \cite{infogeo}.

\section{Time-correlated Markov events}  

	The independent, identically distributed (i.i.d.) time series corresponds to the ideal solution in thermochemistry.  A time series with temporal correlations can be modelled as a Markov process.  We now consider an aperiodic, irreducible Markov chain with $K$ discrete possible states and a transition probability matrix $(p_{ij})_{K\times K}$.  Let us denote its invariant probability as $\{\pi_i\}$: 
\begin{equation}
         \sum_{i=1}^K \pi_ip_{ij} = \pi_j.
\end{equation} 
The LDT for the empirical relative counting frequencies $\{\nu_i\}$ indicates the existence of a hidden, non-normalized measure $\{\xi_i\}$ that accounts for the deviation of $\{\nu_i\}$ away from $\{\pi_i\}$, and gives the LDRF \cite{natarajan,dembo}:
\begin{equation}
   \varphi\big( \{\nu_j\}\big) = \sup_{\{\xi_i\}} \left\{
        -\sum_{j=1}^K  \nu_j \log\left( \frac{1}{\xi_j}
          \sum_{i=1}^K \xi_ip_{ij} \right)
        \right\}.       
\label{ldrf4mc}
\end{equation}
Eq. \ref{ldrf4mc} shows the central role of the matrix 
$\xi_ip_{ij}/\xi_j$, which has a principal eigenvalue $1$ with
$(\pi_1/\xi_1,\cdots,\pi_K/\xi_K)$ and $(\xi_1,\cdots,\xi_K)^T$ as the corresponding left and right eigenvectors.  It defines a conjugate Markov chain with transition probability matrix \cite{cyq-21}
\begin{equation}
\label{cMC}
       \hat{p}_{ij} = \frac{\xi_jp_{ji} }{\xi_i u_i}, \text{ where }
           u_i \equiv \sum_{j=1}^K   \frac{\xi_jp_{ji} }{\xi_i},
\end{equation}
whose invariant probability distribution is the $\{\nu_i\}$ in Eq. \ref{ldrf4mc}.  The $\varphi$  in (\ref{ldrf4mc}) then acquires the Shannon-Sanov form
\begin{equation}
       \varphi\big( \{\nu_i\}\big)  = \sum_{j=1}^K \nu_j
           \log\left(\frac{\nu_j}{v_j}\right),
\end{equation}
where $v_i=\nu_i/u_i$.  Interestingly, the $\{u_i\}$ and $\{v_i\}$ are actually the two principal eigenvectors of yet another positive matrix, $\xi_jp_{ji}/(\xi_iu_j)$.  Under a hidden, prior measure $\{\xi_i\}$ for observing the states and the Bayesian logic, one thus has a pair of positive matrices  \cite{schrodinger,nagasawa} and their corresponding principal eigenvectors:
\begin{subequations}
\label{equation19}
\begin{eqnarray}
	&& \sum_{i=1}^K \nu_i\left(\frac{\xi_jp_{ji}}{\xi_iu_i}\right)
        = \nu_j,  \  \ 	
         \sum_{j=1}^K \left(\frac{\xi_jp_{ji}}{\xi_iu_i}\right)
        = 1,
\\
    && \sum_{i=1}^K v_i \left(\frac{\xi_jp_{ji}}{\xi_iu_j}\right) = v_j, \  \
 	      \sum_{j=1}^K \left(\frac{\xi_jp_{ji}}{\xi_iu_j}\right) u_j = u_i.
\end{eqnarray}
\end{subequations}
The first one is a Markov matrix; the second one solves two related stochastic problems for observing the Markov chain in time w.r.t. the ``internal time measure'' $\xi_i$ of the $i$th state: The number of visits to the $j$th state per unit of time $v_j/\xi_j$ and the mean dwell time in the $k$th state $(u_k\xi_k)^{-1}$.  The stationary probability $\nu_i$ then is $(u_i\xi_i)\times(v_i/\xi_i)$.  

	The conjugate Markov process has the stationary joint probability for a pair of consecutive states $i$ and $j$ $\nu_i\hat{p}_{ij}=v_iu_i\hat{p}_{ij}=\xi_jp_{ji}v_i/\xi_i$.  It preserves the cycle affinity of the original $p_{ij}$ on each and every cycle \cite{yang-cycle-affinity}.  In other words, cycle affinities on all cycles are invariant under observations and its LDT.

\subsection{Duality and Legendre-Fenchel transform}

	To find the Gibbs-Kirkwood potential function introduced in \cite{cyq-21}, as the conjugate variables to $\{\nu_i\}$, we note that the optimization in Eq. \ref{ldrf4mc} yields a set of $\xi_i^*(\{\nu_\ell\})$ at which
$\partial\varphi(\xi^*)/\partial\xi_j=0$.  We therefore have,
\begin{equation}
  \mu_j = \frac{\partial\varphi}{\partial\nu_j}
   = -\log\left[\frac{1}{\xi^*_j(\{\nu_\ell\})}\sum_{i=1}^K \xi^*_i(\{\nu_\ell\})p_{ij}\right] + \hat{\lambda},
\label{eqn21}
\end{equation}
where the additive $\hat{\lambda}$ is the gauge freedom due to normalization $\sum_{i=1}^K\nu_i=1$ \cite{cyq-21}, and the LFT of $\varphi$ in (\ref{ldrf4mc})
\begin{subequations}
\begin{eqnarray}
	&& \sup_{\{\nu_j\}} \left\{\sum_{i=1}^K \nu_i\mu_i-\varphi\big(
                \{\nu_i\}\big)
                 \right\} = \hat{\lambda}
\\
	&=&  \sum_{j=1}^K  \tilde{\nu}_j \log\left( \frac{e^{\mu_j}}{\xi_j^*(\{\tilde{\nu}_\ell\}) }
          \sum_{i=1}^K \xi_i^*(\{\tilde{\nu}_\ell\})p_{ij} \right),
\end{eqnarray}
\end{subequations}
in which optimal $\tilde{\nu}$'s are functions of $\{\mu_k\}$ through implicit function (\ref{eqn21}).  With the presence of $\hat{\lambda}$, the implicit function is invertible.  Eq. \ref{eqn21} can be rearranged into
\begin{eqnarray}
\label{eqn23}
   \sum_{i=1}^K \xi^*_i\big(\{\tilde{\nu}_\ell\}\big) 
     \Big( p_{ij}e^{\mu_j} -  \lambda \delta_{ij}\Big) = 0,
\\
    \text{ or } \   \sum_{i=1}^K
      \frac{ \xi^*_ip_{ij} }{ \xi^*_j} = \lambda e^{-\mu_j},
\label{eqn24}
\end{eqnarray}
where $\lambda = e^{\hat{\lambda}}$. Eq. \ref{eqn23} indicates that  the $\lambda$ is the principal eigenvalue of the matrix $\big(p_{ij}e^{\mu_j}\big)_{K\times K}$, with the corresponding eigenvector $\{\xi^*_i\}$.  Eq. \ref{eqn24} shows that 
$\lambda e^{-\mu_i}= u_i$ in (\ref{cMC}).

	Ellis-G\"{a}rtner theorem says that the LFT of the LDRF $\varphi$ is the limit of a scaled cumulant generating function, which is the principal eigenvalue of the $K\times K$ matrix $\{p_{ij}e^{\mu_j}\}$ \cite{dembo}.  Giving a tilting $e^{u_j}$ to a Markov chain \cite{barato}, the corresponding principal eigenvector provides the hidden measure $\{\xi_i\}$ together with the conjugate Markov chain (\ref{cMC}). Data statistics $\{\nu_i\}\to\{\xi_i\}\to$ $\{\varphi\leftrightarrow\hat{\lambda}\}$ is the logic thread following probabilistic reasoning; internal energy $\{\mu_j\}\to\{\lambda,\xi_j\}\to\{\nu_j=u_jv_j\}$ has been the causal chain of physics.

\subsection{Markov chain with detailed balance}

	If the Markov chain has detailed balance $\pi_ip_{ij}=\pi_jp_{ji}$, then the $u$'s and $v$'s in (\ref{equation19}b) can be explicitly solved in terms of the $\pi$'s and $\xi$'s.  Since the matrix in (\ref{equation19}b)
\begin{equation}
   \frac{\xi_jp_{ji} }{u_j\xi_i} =   \frac{\xi_jp_{ji}/\xi_i }{
        \sum_{k=1}^K \xi_kp_{kj}/\xi_j } =
          \frac{\xi_j^2\pi_ip_{ij}/(\pi_j\xi_i) }{
        \sum_{k=1}^K \xi_kp_{kj} },
\end{equation}
we can verify that its principal left eigenvector is $v_i=\xi_i^2/\pi_i$.  This yields $u_j = \nu_j\pi_j/\xi_j^2$, and the transition probability of the conjugate Markov chain in (\ref{equation19}a):
\begin{subequations}
\label{equation21}
\begin{equation}
          \hat{p}_{ij} = \frac{\xi_jp_{ji}}{\xi_iu_i}
       = \frac{\xi_i\xi_jp_{ji}}{\nu_i\pi_i},
\end{equation}
and the corresponding joint probability for a pair that occurs in the empirical counting data:
\begin{equation}
         \nu_i\hat{p}_{ij} = \frac{\xi_i\xi_jp_{ji}}{\pi_i} =
         \nu_j\hat{p}_{ji}.
\end{equation}
\end{subequations}
Eq. \ref{equation21} provides a system of nonlinear equations to solve the hidden measure $\{\xi_i\}$.  It is also a linear relation that involves the symmetric matrix $\xi_jp_{ji}\xi_i/\pi_i$.

\subsection{Continuous-time Markov process}

	For an infinitesimal $\Delta t$ time interval, the transition probability matrix $p_{ij} = \delta_{ij}+q_{ij}\Delta t$, where $q_{ij}$, when $i\neq j$, is the transition rate from state $i$ to $j$, and $q_{ii}=-\sum_{j\neq i}q_{ij}$.  Eq. \ref{ldrf4mc} becomes \cite{baldi} 
\begin{equation}
        \hat{\varphi}(\{\nu_i\}) = 
 \frac{\varphi( \{\nu_j\}) }{\Delta t} = \sup_{\{\xi_i\}} \left\{
        -\sum_{i,j=1}^K  \nu_j \left[\frac{ \xi_ip_{ij}}{\xi_j}\right]
           \right\},
\label{ldrf4ctmc-0}
\end{equation}
and Eq. \ref{equation21} becomes $\xi_i^2 = \nu_i\pi_i$.  The LDRF in (\ref{ldrf4ctmc-0}) thus is reduced to 
\begin{equation}
  \hat{\varphi}\big(\{\nu_i\}\big) = -\sum_{i,j=1}^K 
      \sqrt{\nu_i} \left(\frac{ \sqrt{\pi_i } q_{ij}}{ \sqrt{\pi_j} } \right)\sqrt{\nu_j},
\label{ldrf4ctmc}
\end{equation}
whose LFT is
\begin{eqnarray}
	&& \sup_{\{\nu_i\}} \left\{ \sum_{i=1}^K \mu_i\nu_i-\hat{\varphi}\big(\{\nu_i\}\big)\right\} 
\nonumber\\
	&=&  \sup_{\{\nu_i\}} \left\{ \sum_{i,j=1}^K 
      \sqrt{\nu_i} \left(\frac{ \sqrt{\pi_i } q_{ij}}{ \sqrt{\pi_j} }+\mu_i\delta_{ij} \right)\sqrt{\nu_j} \right\}
\nonumber\\
	&=& \sup_{\{x_i\}}  \left\{ \frac{\sum_{i,j=1}^K x_i\Gamma_{ij}x_j}{ \sum_{i=1}^K x_i^2 } \right\} = 
         \lambda_{\text{max}}\big[ \vGamma\big(\{\mu_k\}\big) \big],
\end{eqnarray}
in which symmetric matrix $\vGamma$ has elements
\begin{equation}
            \Gamma_{ij}\big(\{\mu_k\}\big) 
          = \left(\frac{ \sqrt{\pi_i } q_{ij}}{ \sqrt{\pi_j} } \right)+ \mu_j\delta_{ij}.
\end{equation}
The conjugate variables to $\nu_k$ is 
\begin{equation}
\label{eqn31}
        \mu_k = \frac{\partial\hat{\varphi} }{\partial\nu_k} = 
           -\sum_{i=1}^K  \sqrt{\frac{\nu_i}{\nu_k} } \left(\frac{ \sqrt{\pi_i } q_{ik}}{ \sqrt{\pi_k} } \right) + \lambda.
\end{equation}
The inverse, $\{\nu_i\}$ as functions of $\{\mu_j\}$, is obtained as an eigenvalue problem:
\begin{equation}
         \sum_{i=1}^K \sqrt{\nu_i} \left[
            \left(\frac{ \sqrt{\pi_i } q_{ik}}{ \sqrt{\pi_k} } \right)  
        + \mu_k\delta_{ik} \right] = \lambda\sqrt{\nu_k}.
\label{eqn26}
\end{equation}
Eq. \ref{eqn26} shows that the gauge freedom $\lambda$ in (\ref{eqn31}) is the principal eigenvalue of matrix $\vGamma$, with
the corresponding eigenvector being $\{\sqrt{\nu}_i\}$.  The $\lambda_{\text{max}}(\{\mu_k\})$ is the LFT of $\hat{\varphi}(\{\nu_i\})$, and vice versa.

	Finally, we have a dynamic equation for the entropy function following the ``thermodynamics of time and events'', {\em e.g.} Eq. \ref{HJE},
\begin{equation}
	\frac{\partial}{\partial t}\Phi\big(t,\{t\nu_i\}\big)  = \lambda_{\text{max}}\left(\left\{ 
          \frac{\partial\Phi}{\partial (t\nu_i) }\right\}\right).
\label{e33}
\end{equation}
The rhs of (\ref{e33}), a Hamiltonian function in terms of conjugate variables $\{\mu_i\}$, is the principal eigenvalue of the symmetric matrix $\vGamma$ which has a Markovian part $\sqrt{\pi_i}p_{ij}/\sqrt{\pi_j}$ and a diagonal Gibbs-Kirkwood energy part $\mu_i\delta_{ij}$.

\subsection{A generic Hamiltonian for continuous-time Markov process }
\label{sec:3d}
The LDRF for both empirical $\{\nu_i\}$ and mean transition rates $\{r_{ij}\}_{i\neq j}$ along a continuous Markov sample trajectory is:
\begin{equation}
    I\big(\{r_{ij}\},\{\nu_i\}\big)
     = \sum_{i,j=1,i\neq j}^K r_{ij}\left(
     \log\frac{r_{ij}}{\nu_iq_{ij}} - 1 + 
      \frac{\nu_iq_{ij}}{r_{ij}}\right),
\label{equa23}
\end{equation}
which is also called level 2.5 LDT \cite{maes,barato}, in which
\[
  r_{ij} = \frac{\# \text{ of } i \to j 
  \text{ transitions }}{\text{total time } t},
\]
and $\nu_i$ is again the fraction of the occupation time in state $i$ within the total time $t$.  The $\{r_{ij}\}$ are constrained under a shift invariance \cite{dembo} $\sum_{j=1,j\neq k} r_{kj}= \sum_{i=1,i\neq k} r_{ik}$, and $\{\nu_i\}$ with a normalization $\sum_{i=1}\nu_i=1$.  Introducing corresponding conjugate variables:
\begin{eqnarray}
 \log v_{k\ell} &=& \frac{\partial I}{\partial r_{k\ell} }+\log\xi_k-\log\xi_{\ell} \ =\ \log\left(\frac{\xi_kr_{k\ell}}{\nu_kq_{k\ell}\xi_{\ell} }\right), \hspace{0.7cm}
\label{equa24}
\\[6pt]
 \mu_k &=& \frac{\partial I}{\partial\nu_k} + \lambda
   \ =\  \lambda - q_{kk} -\sum_{j=1,j\neq k}^K \frac{r_{kj}}{\nu_k},
\label{equa25}
\end{eqnarray}
where $\xi$'s and $\lambda$ are the Lagrangian multipliers for the constraints.  Since the $I$ function in (\ref{equa23}) is locally Ehd-1,  combining the two set of relations in (\ref{equa24}) and (\ref{equa25}) yields yet another EoS in terms of all the conjugate variables and Langrangian multipliers, e.g., a Hamiltonian function:
\begin{subequations}
\label{equa26}
\begin{equation}
   \sum_{j=1,j\neq i}^K \left(\frac{q_{ij}v_{ij}\xi_j}{\xi_i}\right) 
       +  \mu_i+q_{ii}  = \lambda,
\end{equation} 
in which $\xi$'s satisfy the equation
\begin{equation}
     \xi_k\sum_{i=1,i\neq k}^K 
     \Big( \nu_iv_{ik}q_{ik}\Big)\frac{1 }{
      \xi_i} = \frac{1}{\xi_k} \sum_{j=1,j\neq k}^K \Big(\nu_kv_{kj}q_{kj} \Big)\xi_j.
\end{equation}
\end{subequations}
We note Eq. (\ref{eqn31}) is a special case of (\ref{equa26}a) as $v_{ij}=1$, when there is no empirical information on $\{r_{ij}\}$. 

	Re-arranging the two equations in (\ref{equa26}) gives us
\begin{subequations}
\label{HQEVP}
\begin{eqnarray}
    \sum_{j=1,j\neq i}^K \big(q_{ij}v_{ij}\big)\xi_j +
      \big(q_{ii}+\mu_i\big)\xi_i &=& \lambda \xi_i,
\\
    \sum_{i=1,i\neq k}^K \frac{\nu_i}{\xi_i}\big(q_{ik}v_{ik}\big) +
      \frac{\nu_k}{\xi_k}\big(q_{kk}+\mu_k\big)  &=& \lambda\left(\frac{\nu_k}{\xi_k}\right). \hspace{1cm}
\end{eqnarray}
\end{subequations}
Thus the $\lambda$ again has dual properties: First, it is the principal eigenvalue of 
the tilted matrix,
\begin{widetext}
\begin{equation}
\label{TheMatrix}
    \left(\begin{array}{ccccc}
     q_{11}+\mu_1 &  q_{12}v_{12} & q_{13}v_{13} & \cdots & q_{1K}v_{1K} \\
     q_{21}v_{21} & q_{22}+\mu_2 & q_{23}v_{23} & \cdots & q_{2K}v_{2K} \\
     \vdots &  \vdots & \ddots & \cdots & \vdots \\
     q_{K-1,1}v_{K-1,1} &  q_{K-1,2}v_{K-1,2} & \cdots &  q_{K-1,K-1}+\mu_{K-1} & q_{K-1,K}v_{K-1,K} \\
     q_{K1}v_{K1} &  q_{K2}v_{K2} & \cdots & q_{K,K-1}v_{K,K-1} &  q_{KK}+\mu_K
     \end{array}\right),
\end{equation}
\end{widetext}
with $\{\xi_i\}$ and $\{\nu_i/\xi_i\}$ being the corresponding right and left eigenvectors. Their product yields $\{\nu_i\}$, the observed relative frequency of the state.  Second, as a function of all the conjugate variables, it is actually the Legendre-Fenchel transform of function $I$ \cite{barato}: $\lambda\big(\{v_{ij}\},\{\mu_i\}\big) =$
\begin{eqnarray}
      \sup_{\{r_{ij}\},\{\nu_i\}}
         \left\{ \sum_{i,j=1,i\neq j}^K
           r_{ij}\log v_{ij} + \sum_{i=1}^K \mu_i\nu_i 
           - I\big(\{r_{ij}\},\{\nu_i\}\big) \right\}.
\nonumber\\
\end{eqnarray}
In terms of the conjugate variables, an off-diagonal $q_{ij}$ simply serves as a scaling for $v_{ij}$, and a diagonal $q_{ii}$ serves as a reference value for $\mu_i$.  The statistics of the rate of changes is captured by the $\lambda$, as a Hamiltonian function.

	When the $v_{ij}=1$ in the matrix in (\ref{TheMatrix}), we recover the Eq. \ref{equation19}. A contraction \cite{barato} sets a conjugate variable to zero in the duality formalism.  In  thermodynamics, a conjugate variable $Y=\rd I/\rd y$ is understood as the entropic ``thermodynamic force'' that is responsible for the ``changes in $y$".  A thermodynamic equilibrium is when there is no information on a statistical quantity $y$ and thus it is assumed to be at its expected value, e.g., the state of ``maximum entropy'', when the entropic force $X=0$.

\section{Nanothermodynamic treatment}
\label{sec:iv}

	Let us revisit the entropy function for the Poisson process, the $\Phi$ in Eq. \ref{equation16} that includes the sub-extensive $[\cdots]$ term in (\ref{equation16}b).  Corresponding to $t$ and $n$, the conjugate variables $\eta$ and $\mu$ now become:
\begin{subequations}
\label{equation22}
\begin{eqnarray}
	&& \eta = \left(\frac{\partial\Phi}{\partial t}\right)_n 
           = \frac{n}{t}-\lambda,
\\
	 && \mu = -\left(\frac{\partial\Phi}{\partial n}\right)_t = 
         \log \frac{\sqrt{n(n+1)} }{r t}.
\end{eqnarray}
Solving $t$ and $n$ in terms of $\eta$ and $\mu$ from the two equations, we have
\begin{equation}
	 t = \frac{\lambda+\eta}{\lambda^2e^{2\mu}-(\lambda+\eta)^2}, \
          n = \frac{(\lambda+\eta)^2}{\lambda^2e^{2\mu}-(\lambda+\eta)^2}. 
\end{equation}
\end{subequations}
Then the bivariate LFT
\begin{subequations}
\label{nano-biv-lft}
\begin{eqnarray}
	\Psi(\eta,\mu) &=& -\eta t +\mu n - \Phi(t,n)
\nonumber\\
     &=&\frac{1}{2}\log\left(\frac{\lambda^2e^{2\mu}-(\lambda+\eta)^2}{  \lambda^2e^{2\mu} } \right),
\\
	 \rd\Psi(\eta,\mu) &=&          
      - \frac{(\lambda+\eta)\rd \eta}{\lambda^2e^{2\mu}-(\lambda+\eta)^2}
     + \frac{(\lambda+\eta)^2\rd\mu}{\lambda^2e^{2\mu}-(\lambda+\eta)^2} 
\nonumber\\
	&=&  -t\rd\eta + n\rd\mu. 
\end{eqnarray}
\end{subequations}
Eq. \ref{nano-biv-lft}b has the form of the Hill-Gibbs-Duhem equation \cite{lu-qian-20}.  The $\Psi/t\sim o(1)$ as $t, n\to\infty$, yields a singular LFT of $\varphi$ in Eq. \ref{eos}.  This can be best seen from 
the two Hessian matrices of functions $\Phi(t,n)$ and $\Psi(-\eta,\mu)$, are:
\begin{equation}
    \frac{\mathscr{D}[-\eta,\mu]}{\mathscr{D}[t,n]} = \left(\begin{array}{cc}
        \frac{n}{t^2} & -\frac{1}{t} \\[5pt]   
              -\frac{1}{t}  &  \frac{2n+1}{2n(n+1)}
            \end{array}\right) 
\label{2matrices-a}
\end{equation}
and 
\begin{eqnarray}
     \frac{\mathscr{D}[t,n]}{\mathscr{D}[-\eta,\mu]}
      &=&  \left(\begin{array}{cc}
     -\frac{\lambda^2e^{2\mu}+(\lambda+\eta)^2}{[\lambda^2e^{2\mu}-(\lambda+\eta)^2]^2} &  
        -\frac{2(\lambda+\eta)\lambda^2 e^{2\mu} }{[\lambda^2e^{2\mu}-(\lambda+\eta)^2]^2} \\[5pt]
         -\frac{2(\lambda+\eta)\lambda^2 e^{2\mu} }{[\lambda^2e^{2\mu}-(\lambda+\eta)^2]^2}     &   -\frac{2(\lambda+\eta)^2\lambda^2 e^{2\mu}  }{[\lambda^2e^{2\mu}-(\lambda+\eta)^2]^2}     \end{array} \right) 
\nonumber\\[5pt]
      &=& \left(\begin{array}{cc}
        -\frac{(1+2n)t^2}{n} & -2(1+n)t \\[5pt]   
              -2(1+n)t  &  -2n(1+n)
            \end{array}\right) .
\label{2matrices-b}
\end{eqnarray}
The duality between $\Phi(t,m)$ and $\Psi(\eta,\mu)$ implies the Hessian matrix of the former is the inverse Hessian matrix of latter, with one-to-one correspondence between $(\eta,\mu)$ and $(t,n)$ via Eq. \ref{equation22}.  

These results return to those in the previous section as $n+1\simeq n$, or $\lambda e^{\mu}\simeq \lambda+\eta$, in Eq. \ref{equation22}.  In this case, both $t, n\to\infty$ in (\ref{equation22}c), and their ratio $\frac{n}{t}=r e^{\mu}$.
We observe that the Hessian matrix of $\Phi(t,n)$ given in (\ref{2matrices-a}) becomes singular, with one of the two eigenvalues becoming zero.  The singularity gives rise to the EoS.

	In Hill's nanothermodynamics \cite{hill-book}, the $\Psi$ is called a subdivision potential.  Following his work, one further has the integral ``chemical potential'' for an event, $\hat{\mu}$,
\begin{equation}
    \hat{\mu} =  \frac{\Phi}{n} =
   \log\frac{\sqrt{n(n+1)}}{\lambda t} - 1 + \frac{\lambda t}{n}  +  \frac{\log (n+1)}{2n},
\end{equation}
which is related to the differential $\mu$ in (\ref{equation22}b):
\begin{equation}
 \mu -\hat{\mu} = 1-
   \frac{\lambda t}{n} - \frac{\log (n+1)}{2n}
        = n\left(\frac{\partial\hat{\mu}}{\partial 
       n}\right)_t.
\end{equation}
This is a set of novel thermodynamic relations that only exist for small systems.

	Curiously, we note that the determinant of (\ref{2matrices-a}) is
\[
              \frac{1}{t^2}\left(\frac{2n+1}{2(n+1)}-1\right)
        = -\frac{1}{2(n+1)t^2} < 0,
\]
implying the ``entropy function'' for time and events, $\Phi(t,n)$, is not necessarily convex.

\subsection{Singular transformation and subadditive potentials}

	As $t,n\to\infty$ with the same order, the conjugate variables in (\ref{equation17}) are $\sim O(1)$.  More importantly, the change of
variables $(t,n)\to (\eta_0,\mu_0)$ is non-invertible.  The $\eta$ and $\mu$ in (\ref{equation22}) differ from those in (\ref{equation17}) by a term $\sim o(1)$, this makes $(t,n)\to (\eta,\mu)$ invertible.  

	If one denotes
\[
    \Phi_1(t,n) = -\frac{n+1}{2}\log(n+1)-\frac{n}{2}\log n,
\]
then one has an interesting observation:
\begin{eqnarray}
	\Psi(\eta,\mu) &=& -\eta t + \mu n - \Phi(n,t)
\nonumber\\
	        &=& -(\eta-\eta_0)t + (\mu-\mu_0)n - \Phi_1(n,t)
\nonumber\\
	       &=& -\eta_1 t + \mu_1 n - \Phi_1(n,t),
\end{eqnarray}
in which
\begin{equation}
		\eta_1 = \left(\frac{\partial\Phi_1}{\partial t}\right)_n, \
       \mu_1 = -\left(\frac{\partial\Phi_1}{\partial n}\right)_t.
\end{equation}

	In general, as $t$ and $n=\nu t$ $\to\infty$, one expects that 
\begin{equation}
   \Phi_1(t,n) \sim t^{\kappa}\varepsilon(\nu), \  \kappa<1,
\end{equation}
and
\begin{subequations}
\label{eqn42}
\begin{eqnarray}
		\eta_1 &=& t^{\kappa-1} \big[ \kappa\varepsilon(\nu)-
             \nu\varepsilon'(\nu) \big], 
\\
	\mu_1 &=& -t^{\kappa-1} \varepsilon'(\nu),
\\
	\frac{\eta_1}{\mu_1} &=& \nu  -\kappa\left(\frac{\varepsilon'(\nu)}{
            \varepsilon(\nu)} \right)^{-1}.
\end{eqnarray}
\end{subequations}

In fact, if one can identify the subadditive exponent $\kappa$, then all the entropy analysis can be carried out once more for 
Ehd-1 $t^{-\kappa}\Phi_1(t,n)$.

We note the set of equations in (\ref{eqn42}) will be very different if the subadditive
term is $\sim O(\log t)$.  In this case, $\Phi_1(t,n)\sim (\log t)\varepsilon(\nu)$ and
\begin{subequations}
\begin{eqnarray}
		\eta_1 &=& t^{-1} \big[ \varepsilon(\nu)- (\log t/t)
             \nu\varepsilon'(\nu) \big], 
\\
	\mu_1 &=& -(\log t/t) \varepsilon'(\nu),
\\
	\Psi(\eta_1,\mu_1) &=& 
          -\log t\ \big[ \nu\varepsilon'(\nu)+ \varepsilon(\nu)\big] + O(1).
\end{eqnarray}
\end{subequations}

\subsection{Nanothermodynamic analysis of Markov processes}

All the LDRFs for empirical relative counting frequency, all four functions in Eqs. (\ref{ldrf4mc}), (\ref{ldrf4ctmc-0}), (\ref{ldrf4ctmc}), and (\ref{equa23}) have the normalization constraint $\sum_{i=1}^K\nu_i=1$.  Using the method of Lagrangian multiplier for the constraints optimization, an additive constant $\lambda$ (gauge freedom) arises, and the LFTs of these LDRFs become precisely the 
$\lambda$.  Identifying the LFT of the LDRF of the frequencies of counting along time, the $\lambda$ is identified as a Hamiltonian function in a HJE.

	A nanothermodynamic analysis of generic Markov process is outside the scope of the present work.  The contribution to the next-order term to the LDRFs in Eqs. (\ref{ldrf4mc}), (\ref{ldrf4ctmc-0}), (\ref{ldrf4ctmc}), and (\ref{equa23}) is expected from the the second eigenvalue and the relative contributions of the two eigen modes, which is dictated by the non-stationary initial distribution.  For a finite state Markov chain, this next-order non-extensive term is $\sim O(1)$.  Continuous state space spatial stochastic models with ``surface effect'' will offer much more interesting and challenging subjects for nanothermodynamic analysis.

\section{Discussion}
\label{sec:discuss}

	All results in the previous sections are based on mathematics;
the derived relations, however, require interpretations.  In traditional applied mathematics, interpretations are based on existing narratives within a scientific field.  However, for the present work such a discussion seems to be beyond the current understanding of theoretical physics, on mechanics and on statistical thermodynamics. The most unsettling element of the implication is that the law of mechanics could be understood through an entropic theory.  This idea is not new \cite{verlinde}; the verbiage below, therefore, could be considered as a part of a scientific hypothesis.  In terms of a mathematical limit, there are several widely used idealizations in our understanding of the physical world: inifitely large systems and infinitely long time stationary processes, for examples.  The present work explores the statistics of infinitely divisible time itself.

{\bf\em Newtonian mechanical energy and the Gibbs-Kirkwood energy.}
The Gibbs-Kirkwood energy introduced in \cite{cyq-21} is a statistical concept.  Let us put appropriate units of physically measurable quantities from the real world into the discussion:  If one identifies $t$ with time, and its conjugate variable with ``mechanical energy'', then their product has the same dimension as $\hbar$, the reduced Planck constant. One thus could have the asymptotic probability distribution for the events $\asymp e^{-(t/\hbar)\varphi(\mu)}$, where $t\varphi(\mu)/\hbar\equiv \tilde{\Psi}_2(t,\mu)$ is necessarily an Ehd-1 function of variable $t$.  Therefore $H= \varphi(\mu)$, as an EoS, can be interpreted as the Newtonian ``mechanical energy function'' or Hamiltonian, the conjugate variable to time $t$, under the asymptotic limit of $(t/\hbar)\to\infty$. 

	Could the two terms in the symmetric matrix $\vGamma$ in Eq. \ref{eqn26}, $\Gamma_{ij} = \sqrt{\pi_i}p_{ij}/\sqrt{\pi_j} + \mu_i\delta_{ij}$, be interpreted as the {\em kinetic energy} and 
{\em potential energy} in a Schr\"{o}dinger operator?  We note that since the $\vGamma$ is a linear operator to which the eigenvectors are 
$\{\sqrt{\nu_i}\}$, linear superposition of two vectors can lead to 
an interference pattern in probability \cite{dirac}:
\[
        \left(\sqrt{\nu^{(1)}_i}+\sqrt{\nu^{(2)}_i } \right)^2
= \nu^{(1)}_i + \nu^{(2)}_i + 2\sqrt{\nu^{(1)}_i}\sqrt{\nu^{(2)}_i}.
\]
The phenomenon we call {\em wave} is in essence ``counting in space and time''.  In quantum physics, one has a {\em material wave} with Einstein's relation $E=\hbar\omega_{mw}$ and de Broglie relation $Mv=2\pi\hbar\lambda_{mw}$, where $E$ and $Mv$ are energy and momentum of a particle with mass $M$, $\omega_{mw}$ and $\lambda_{mw}$ are the angular frequency and wavelength, and $\hbar=1.055\times 10^{-34}$ m$^2\cdot$kg$\cdot$sec$^{-1}$ is reduced Planck's constant. 

	In Gibbs' chemical thermodynamics, one identifies the conjugate variable of $n$, the counting number of particles, as Gibbs' chemical potential. Then $\mu =-\tilde{\Psi}_1(t,\eta)/t\equiv -\psi_1(\eta)$ can be interpreted as the Gibbs-Kirkwood potential energy. The ``large number'' here seems to be the inverse of Boltzmann's constant, $k_B=
1.381\times 10^{-23}$m$^2\cdot$kg$\cdot$sec$^{-2}\cdot$K$^{-1}$, the unit for entropy in thermodynamics.

	{\em Time} and {\em counting events} are two most fundamental concepts.  With respect to $\hbar$, in the limit of a long time  $[0, t]$ with a large number of events $n$, counting the number of Poissonian events $n$ is an accurate measure of time $t=n/r$, where $r$ is the rate. When $t\neq n/r$, function $\Phi_0(t,n)$ gives the free energy (FE), or ``information'', of a ``Universe'' associated with the deviation.  Eq. (\ref{equation18}) then suggests that $\eta$ is the marginal decrease in FE per unit time, e.g., energy dissipation rate, and $\mu$ is the marginal increase of FE per event, e.g., its chemical potential.  

\begin{quote}
	{\em What is time?  Time is the accumulation of events, and events contain information.  A scientific time is quantified by statistical correlations through the $\{\xi_i\}$ which has contributions from both prior $\{\pi_i\}$
and posterior $\{\nu_i\}$. 
}
\end{quote}

{\bf\em Gauge freedom and conservation laws.}
The appearance of $\hat{\lambda}$ and $\lambda$ in Eqs. \ref{eqn21}) and \ref{equa26} elicits another interpretation in one's sense of theoretical physics:
\begin{equation}
 \sum_{j=1,j\neq i}^K \left(\frac{q_{ij}v_{ij}\xi_j}{\xi_i}\right) 
       + \Big[ \mu_i+q_{ii}  \Big] = \lambda.
\label{EQ45}
\end{equation}
in which we could identify the first term as a ``kinetic energy'' associated with the $i\to j$ transition and the second term in $[\cdots]$ as the Gibbs-Kirkwood potential energy associated with state $i$.
Both quantities change along a trajectory but their sum remains a constant $\lambda$.  By this interpretation, the Lagrangian multipliers in connection with the Shannon-Sanov relative entropy satisfy, as derived in \cite{cyq-21},
\begin{equation}
   \log\left(\frac{\nu_j}{\pi_j}\right) + \mu_j = \lambda.
\end{equation}
This indicates that the ``potential function within an arbitrary additive constant'' can be viewed as a limiting case of (\ref{eqn21}) with
$p_{ij}=\pi_j$, when the correlation between consecutive measurements with low time resolution has fully delayed; the contribution from the kinetic part disappears. The $\lambda$ now becomes the Helmholtz free energy serving as an absolute reference point for all $j$'s.  The additional Langrangian multipliers in connection with mean transition rates $r_{ij}$'s satisfy Eq. \ref{equa26}b.

{\bf\em The Legendre transform.}
When carrying out the optimization in a Legendre-Fenchel transform  in terms of calculus, Legendre transform of a function $\phi(x)$ yields $\psi(y) = x(y)y-\phi(x(y))$ where $x(y)$ is solved from implicit function $y=\phi'(x)$.  If $\phi(x)$ is convex, then the inverse function of $y=\phi'(x)$ exists and is unique.

Legendre transform of an Ehd-1 $\phi$ should be understood as {\em obtaining a nonlinear differential equation for the function $\phi$ in terms of all its partial derivatives}. It introduces a gauge freedom.  There is a deep relation between the Legendre transform and Noether's theorem \cite{jherman}: The HJE such as (\ref{hje0}) can be viewed as an equation for differential operators.  The $y$ variables could be understood as ``generators'' in the language of Lie group and Lie algebra, as well as a ``tangent vector'' in the language of affine geometry.

\section*{Acknowledgement}
  
I thank Erin Angelini, Jeffrey Commons, Wenqing Hu,
Zhiyue Lu, Lowell Thompson, and Ying-Jen Yang for many help discussions.  I am grateful to Profs.
Jin Feng (U. Kansas), Hao Ge (Peking U.), Chun Liu (IIT, Chicago), Quanhui Liu (Hunan U.), Xiang Tang (Wash. U.), Bai-Ling Wang (ANU), and Hong Zhao (Xiamen U.) for continuous advices.

\appendix

\section{Duality theory of  level 2.5 LDRF for diffusion process}

	If the duality analysis of Shannon-Sanov LDRF for empirical i.i.d. sample frequency yields the very Gibbs' theory of ensemble tegether with a novel statistical interpretation for internal energy \cite{cyq-21}, and a duality analysis of the LDRF in (\ref{equa23}) suggested a ``mechanical'' energy conservation, what would be the possible implication from a duality analysis in parallel to Sec. \ref{sec:3d} above for a diffusion process in $\mathbb{R}^d$?  In this case, Eq. \ref{equa23} becomes \cite{barato}:
\begin{equation}
\label{equa30}
    I\big[r,\nu\big] = \left\{\begin{array}{ll}
      \displaystyle 
      \frac{1}{4}\int 
    \Big[\big(r(x)-J_{\nu}(x)\big)\nu^{-1}(x) \times \\
      \hspace{15pt} D^{-1}(x) \big(r(x)-J_{\nu}(x)\big)
    \Big]\rd x  & \nabla\cdot r(x) = 0,
\\
     \infty  & \text{ otherwise}
    \end{array} \right.
\end{equation}
in which $r(x)$ is a zero-divergence empirical flow field, $\nu(x)$ is an empirical occupation frequency, and $J_{\nu}(x) \equiv -D(x)\nabla\nu(x) + b(x)\nu(x)$ denoting the expected stationary flow corresponding to $\nu(x)$.  

	We now consider the duality problem to (\ref{equa30}).  Let us denote the conjugate variables, vector-valued function $v(x)=\delta I/\delta r$ and scalar function $u(x)=\delta I/\delta\nu$. Then under the constraints $\nabla\cdot r(x)=0$ and normalization of $\nu(x)$, calculus of variation yields two relations:
\begin{eqnarray}
     \tilde{v}(x) &=& \frac{1}{2} \Big(\nu(x)D(x)\Big)^{-1}\Big( r(x)-J_{\nu}(x)\Big),
\\
   \tilde{u}(x) &=& -\frac{1}{2}\ b(x)\cdot\Big(\nu(x)D(x)\Big)^{-1}
   \Big(r(x)-J_{\nu}(x)\Big) \hspace{0.7cm}
\nonumber\\
   &-& \frac{1}{2} \nabla\cdot\Big[\nu^{-1}(x) \Big(r(x)-J_{\nu}(x)\Big) \Big]
\\
  &-& \frac{1}{4} \Big(r(x)-J_{\nu}(x)\Big)
\Big(\nu^2 D(x) \Big)^{-1}\Big(r(x)-J_{\nu}(x)\Big),
\nonumber
\end{eqnarray}
in which $\tilde{v}(x)=v(x)+\nabla\xi(x)$ and $\tilde{u}(x)=u(x)-\lambda$. The gauge freedoms $\xi(x)$ and $\lambda$ are the Lagrangian multipliers corresponding to the constaints. They give rise to a pair of first-order partial differential relations between $\big(r,\nu\big)(x)$ and  $\big(\tilde{v},\tilde{u}\big)(x)$, where $\nu(x),\tilde{u}(x)$ are scalar functions and $r(x),\tilde{v}(x)$ are tangent vector fields:
\begin{subequations}
\label{A4}
\begin{eqnarray}
  D(x)\nabla\nu(x) - \Big( b(x)
   + 2D(x)\tilde{v}(x) \Big) \nu(x) &=& -r(x), 
\nonumber\\[-4pt]
\\[-4pt]
  \nabla\cdot\big(D(x)\tilde{v}(x)\big)+
    \Big( b(x) 
      + D(x)\tilde{v}(x) \Big)\cdot \tilde{v}(x)  &=& -\tilde{u}(x).
\nonumber\\[-4pt]
\end{eqnarray}
\end{subequations}
If we assume that $\tilde{v}(x) =\nabla\log\psi(x)$ is a gradient vector field with a potential function, then Eq. \ref{A4}b can be transformed into $(\mathscr{L} + u) \psi(x) = \lambda \psi(x)$, where $\mathscr{L}=\nabla\cdot D(x)\nabla + b(x)\nabla$ is the generator for the diffusion process.  Since (\ref{equa30}) is locally Ehd-1 w.r.t. variables $r(x)$ and $\nu(x)$, one expects the normalization of $\nu(x)$ yields $\lambda[(u,v)(x)]$ as the LFT of $I[(r,\nu)(x)]$, and thus a Hamiltonian function for mechanical systems represented by empirical $r(x)$ and $\nu(x)$, or equivalently the ``statistical mechanical ensemble'' characterized by $(\tilde{v},\tilde{u})(x)$.

	The contraction of $I[r,\nu]$ to Donsker-Varadham's variational formula $I_{DV}[\nu]$ \cite{barato} corresponds to setting vector field $v(x)=0$ in the duality formalism.  A duality analysis of $I_{DV}[\nu]$ yields
$u=\delta I_{DV}/\delta\nu$, and 
\begin{subequations}
\label{A5}
\begin{eqnarray}
      \big( \mathscr{L} + u\big)[\xi] &=&  \lambda \xi,
\\
     \big(\mathscr{L}^* + u \big)\left[\frac{\nu}{\xi}\right]
  &=& \lambda\left(\frac{\nu}{\xi}\right),
\end{eqnarray}
\end{subequations}
where $\mathscr{L}=\nabla\cdot D(x)\nabla + b(x)\nabla$ and
$\mathscr{L}^*[w]=\nabla\cdot [ D(x)\nabla w -  b(x) w]$.  This
agrees with Eq. \ref{equation19} and Eq. \ref{HQEVP}.  In lieu of a geometric meaning,  we note that combining (\ref{A5}a) and (\ref{A5}b) by eliminating $u$ yields $(\xi/\nu)\mathscr{L}^*[\nu/\xi]=\xi^{-1}\mathscr{L}[\xi]$.

\end{document}